\documentclass[twocolumn,aps,prl]{revtex4}
\usepackage{graphicx}% Include figure files
\usepackage{dcolumn}% Align table columns on decimal point
\usepackage{bm}% bold math
\begin{document}

%\draft
%\preprint{Manuscript No. 1}
\title{Nonlinear Schr\"{o}dinger Equation for Quantum Computation}

\author{M. Cemal Yalabik}
%\address{Department of Physics, Bilkent University, 06800 Ankara, Turkey}
\affiliation{Department of Physics, Bilkent University, 06800 Ankara, Turkey }

\date{Received ..., Revised July 25, 2003}
%\maketitle

\begin{abstract}
Utilization of a quantum system whose time-development is described by 
the nonlinear Schr\"{o}dinger equation in the transformation
of qubits would make it possible to construct quantum algorithms which would
be useful in a large class of problems. An example of such a system for
implementing the logical NOR operation is demonstrated.
\end{abstract}

%\pacs{PACS numbers: 72.10.-d, 02.70.-c, 02.60.Cb}

\maketitle

Quantum computing algorithms\cite{qc-algo}
make use of the possibility of parallel
operations on states which make up a superposed set. After appropriate
operations have been made, a number of measurements may be carried out,
resulting in the collapse of the wavefunction to a smaller set, from which
information related to the solution of the particular problem may be deduced.

Emphasis in previous work on quantum computing algorithms has been on
time independent
unitary operations on the superposed states. Focus on such
operations is natural, as the superposed states need to be associated with
relatively simple quantum degrees of freedom, whose time development could
be described by a Schr\"{o}dinger equation incorporating a simple Hamiltonian.
On the other hand, it was noted by Abrams and Lloyd\cite{Abrams} that
a much richer set of problems can lend themselves to solution by
quantum computation if nonlinear evolution of the qubit states could
be realized. In particular, they demonstrate that the availability of a special nonlinear
operation on a single qubit enables the construction
of a quantum algorithm whose repetitive application results in efficient 
progressive separation of searched states from others. They also show that 
implementation of a two bit nonlinear quantum AND gate allows for an
algorithm which finds the answers to an NP-complete problem with
certainty in linear time. It is indicated that the two bit nonlinear 
transformation
itself may be obtained through ordinary unitary operations in combination
with single qubit nonlinear operations.

Such nonlinear single qubit operations have been studied\cite{nl-fidelity1,
nl-fidelity2,nl-fidelity3,nl-fidelity4,nl-fidelity5},
with the emphasis
being on the analysis of fidelity in obtaining the results of these
operations through unitary transformations. 

A possibility for realizing nonlinear quantum operations is through a system dynamics described by the
nonlinear Schr\"{o}dinger equation which
appears in the analysis of Bose-Einstein condensation\cite{BEC} (BEC) and
other contexts\cite{ginzburg}. 
Shi\cite{Shi} has demonstrated that nonlinear quantum evolution is 
possible in Bose-Einstein condensates coupled to one another through a 
tunneling junction. An explicitly nonlinear equation of motion for the
containment well occupation coefficients has been derived. An important 
feature of this work is that it demonstrates how entanglement in BEC is
realized, starting from fundamental considerations. The work also contains
a discussion of the possibility of utilizing this nonlinearity in quantum 
computations.

Admittedly, these equations involve
approximations, and appear only because the ``background" in a collective
quantum system is treated in some mean-field form. The discussion of the
validity of the range of the approximation to a particular implementation
may be deferred as a technical detail. However, the utilization of a
(possibly macroscopic) collective quantum event as a qubit raises deeper
questions which must be answered. We briefly touch upon this point in the 
concluding paragraph.

In this paper, we will assume that the dynamics of such nonlinear
Schr\"{o}dinger equations may be applicable to operations on qubits and
look into the possibilities introduced by such operations. 
It will be shown that a two-qubit quantum gate can be constructed with
a single condensate (in contrast to the more than one in \cite{Shi})
in the presence of a non-uniform potential. It will also be pointed out that 
availability of such nonlinear quantum operations allow the efficient
search of an optimal solution through
pairwise elimination of possibilities making up an extended solution set.

A typical
quantum computation algorithm utilizes the creation of a superposition of
parallel states, usually a complete set of enumerable states:
\begin{equation}
 |\Psi_1 > = \frac{1}{\sqrt{N}}\sum_{k=0}^{N-1} |k>  .  \label{eq:psi1} 
\end{equation}

If the qubits forming the pure state $|k>$ is formed by $n$ quantum states
with two possible eigenstates each, then $N=2^{n}$, and $k$ may be taken as
the number corresponding to the binary representation generated by the $n$
qubits. 

The space is then enlarged to include a function of $k$ in the representation:
\begin{equation}
 |\Psi_1> \rightarrow |\Psi_2> = 
   \frac{1}{\sqrt{N}}\sum_{k=0}^{N-1} |k> |f(k)> . \label{eq:psi2} 
\end{equation}

The extended space containing $|f(k)>$ itself is made up of additional
qubits. A measurement carried out on this state will yield a superposition
of a smaller subset:
\begin{equation}
 |\Psi_2> \rightarrow |\Psi_3> = 
   \frac{1}{\sqrt{N'}}\sum_{k}' |k> |f(k)>  \label{eq:psi3} 
\end{equation}
where the prime on the summation indicates a sum over a subset of the
states $|k>$ consistent with the results of the measurement, a total number
of $N'$. Similar operations involving enlargements of the space, unitary
transformations on the superposed states, and measurements may be carried
out until a final set of measurements will yield results relevant to the
solution of the problem at hand. This may be in the form of a direct result
yielding a numerical value\cite{shore}, or one may have a statistical result, in
which a sufficient number of measurements must be repeated to obtain an
average quantity with sufficient accuracy\cite{probabilistic}.

To motivate the utility of nonlinear transformations in quantum computing,
consider the problem of searching through all possible ways of completing a
task, to find an optimal one. The number $k$ will represent one of the
possible pathways, and we will assume that the binary representation of
this number ({\em i.e.} values of the qubits) can be grouped into successive
``moves" or ``choices" which must be carried out to follow this pathway. 
For example, consider the following qubit decomposition of the 
state $|k>$:
\begin{equation}
 |k> = \left | \underbrace{ q_1 q_2 \cdots q_m}_{\rm move\; 1} 
                 \underbrace{q_{m+1} \cdots q_{2m} }_{\rm move\; 2}
                 \underbrace{q_{2m+1}\cdots q_{3m} }_{\rm move\; 3}
  \cdots \right >  \label{eq:decomp} 
\end{equation}

This would correspond to labeling all possible pathways which could be
reached by a finite number of discrete moves to be chosen from a finite
number of possibilities. For example, for the traveling salesman
problem\cite{traveler} with 256
cities, one could assign consecutive $m=8$ qubits to represent the city to be
visited at that stage, and one would need 256 such 8-qubits to represent
the complete trip. (Obviously, this procedure would also produce some
pathways which are ``illegal" in the way this problem is defined, but these
will be discarded in the solution.) Alternatively, if the remaining
consecutive steps in an ongoing chess game were to be described by $|k>$,
then one could use 12 consecutive qubits to describe motion of a piece from
a general point on the $8 \times 8$ square to another point on the square.
All possible games with a total of 100 moves could be represented by a
total of $12 \times 100$ qubits. (Again, this type of coding generates an
overwhelming ratio of illegal moves, which need to be discarded.)
The function $|f(k)>$ is arranged to hold information about the end result
of decision process, for example, whether the completed moves correspond to
a ``legal" sequence, and if so, what the result is. The ``result" here
would be whether the game has been ``won" or, what the total
distance traveled is within the context of the traveling salesman problem
mentioned above.

Note that the superposed state may be factored so that it can be expressed
as a sum over the more significant qubits representing the number $k$,
multiplying the two terms corresponding to the least significant qubit:
\begin{equation}
 \frac{1}{\sqrt{N}}\sum_{j=0}^{N/2-1} |j> ( |0> |f(j|0)> + 
   |1> |f(j|1)>  )       \label{eq:factor} 
\end{equation}
where the notation $(j|0) = 2j$ and $(j|1)=2j+1$ has been used. 
Note that for each $j$, there is a ``preferred" choice between the cases
$(j|0)$ and
$(j|1)$ (based on the values of $f(j|0)$ and $f(j|1)$), which we will label
as $j'$. One could then obtain a superposed state with a reduced number
of terms if the following transformation could be made:
\begin{equation}
\frac{1}{\sqrt{2}}(|0> |f(j|0)> + |1> |f(j|1)> ) \rightarrow |f(j')>
\label{eq:trans}
\end{equation}
so that the new state is
\begin{equation}
\frac{1}{\sqrt{N/2}}\sum_{j=0}^{N/2-1} |j>  |f(j')> .
\end{equation}
This process could then be iterated until a single qubit remains, yielding
its optimal value. Once this value is determined, the problem is reduced to
the determination of the remaining $N-1$ qubits, for which the above
process must be repeated, starting with these $N-1$ qubits. The number of
operations necessary for the 
determination of all of the qubits then can be seen to be proportional to
$N^2$, one factor of $N$ coming from the repetition of the operation for
each qubit to be determined, and another factor from the number of
transformations of the type shown in Eqn.~\ref{eq:trans}.

For many problems of interest however, including the ones mentioned above
as examples, this transformation cannot be achieved with unitary
operations. 
Nonlinearity allows for ``communication" between pairs
of superposed states in carrying out the operation in Eqn.~\ref{eq:trans}.
For example, the simple logical NOR operation (which is 
related to how the
``legality" operation of the moves would transform) would need to have

\begin{eqnarray}
  \frac{1}{\sqrt{2}}(|0> |0> + |1> |0>) & \rightarrow & |1> \label{eq:one} \\
  \frac{1}{\sqrt{2}}(|0> |0> + |1> |1>) & \rightarrow & |0> \label{eq:two} \\
  \frac{1}{\sqrt{2}}(|0> |1> + |1> |0>) & \rightarrow & |0> \label{eq:three} \\
  \frac{1}{\sqrt{2}}(|0> |1> + |1> |1>) & \rightarrow & |0> \label{eq:four} .
\end{eqnarray}

The operation is clearly nonlinear,
as relations~\ref{eq:one}-\ref{eq:four} above demonstrate:
The sum of the left sides of the expressions \ref{eq:one} and
\ref{eq:four} 
of the transformation equals the sums of the left hand sides of the
expressions \ref{eq:two} and \ref{eq:three}, but the same
obviously is not true for the right hand sides.
% Such transformations are possible only through the utilization of nonlinear
% transformations.

A similar discussion may be carried out for optimization problems of the
traveling salesman type, obtaining an optimal set of ``moves" by the
pairwise elimination of non-preferred choices.  Details cannot be provided here,
but it has to be remembered that the standard tools of classical logical
computation (such as the logical AND and OR operations) may be utilized if
nonlinear transformations are to be allowed. This feature then 
(at least in principle) makes
accessible to quantum computation all classical problems which can benefit
from parallelism.

The transformation in expressions \ref{eq:one}-\ref{eq:four} 
may be implemented as the result of
time-development through a nonlinear Hamiltonian. As an example, we will
consider a Hamiltonian of the Gross-Pitaevskii 
type\cite{gross1,gross2,pitaevskii}
in which an extra potential
term proportional to the square magnitude of the wavefunction appears. The
two qubit state $|q_0 q_1>$ with $q_0$ and $q_1$ equal to 0 or 1 will be
taken to be related to the occupation of four sites at 
${\rm \bf r}(q_0,q_1)=(q_0
\hat{\imath} + q_1 \hat{\jmath} )\Delta x$ where $\hat{\imath}$ and
$\hat{\jmath}$
are the unit vectors in the $x$ and $y$ directions respectively. The system
then corresponds to a set of four quantum sites arranged as a square with
side $\Delta x$. The Schr\"{o}dinger Equation describing the system will then be
\begin{eqnarray}
\lefteqn{i\hbar \frac{\partial}{\partial t} \psi(q_0,q_1) =} \nonumber \\
& & -\frac{\hbar^2}{2m(\Delta x)^2} \left [ \psi(\bar{q_0},q_1) + 
  \psi(q_0,\bar{q_1})- 2\psi(q_0,q_1) \right ] + \nonumber \\
& & \left [ V(q_0,q_1) + \alpha |\psi(q_0,q_1)|^2 \right ] \psi(q_0,q_1) 
\end{eqnarray}
where $\alpha$ is a measure of the nonlinearity in the system, and we have
used the shorthand notations $\bar{q}=1-q$ and 
$\psi(q_0,q_1)=\psi({\rm \bf r}(q_0,q_1))$. The first
term on the right hand side is the kinetic energy term of the square geometry,
it may also result
through a tight-binding interpretation of the interaction among four
quantum wells. The coefficient of this term  
$\epsilon = \hbar^2/2m(\Delta x)^2$ with
energy units sets the physical scales of the system. 
The external potential $V$ is the quantity to be
``engineered" so that the time-development of $\psi$ has the required form.

\begin{figure}
	\begin{center}
		\includegraphics{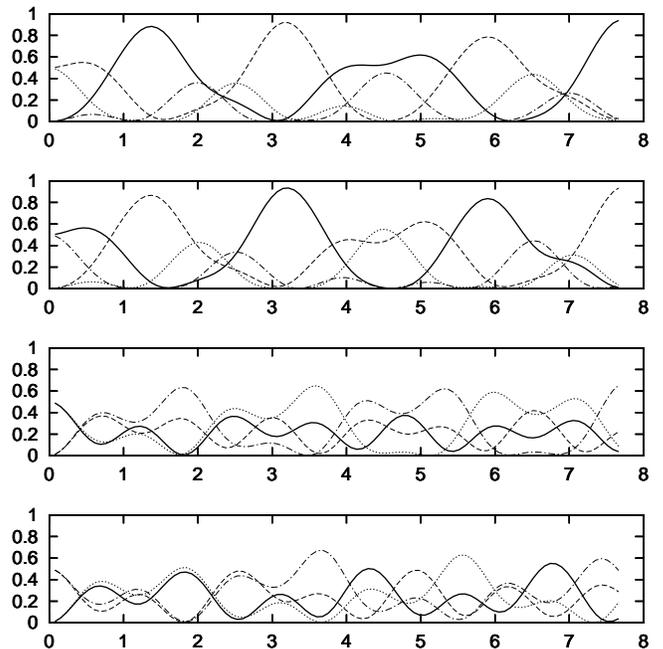}
	\end{center}
%	\caption{tttest}
%	\label{fig:tt}
%\end{figure}

%\begin{figure}
%\includegraphics{tt}% Here is how to import EPS art
\caption{\label{fig:epsart} The squared magnitudes of $\psi(0,0)$ 
(dashed lines),
$\psi(0,1)$ (dotted lines), $\psi(1,0)$ (dot-dashed lines), and $\psi(1,1)$ 
(thick continuous lines) for four different initial conditions. The initial
conditions correspond to the left hand terms of expressions \ref{eq:one}
through \ref{eq:four} from top to down respectively. Parameters of the
numerical computation are given in the text. Note that the final value of
square 
magnitude of $\psi(1,1)$ correlates with the right hand terms of expressions
\ref{eq:one} through \ref{eq:four}.}
\end{figure}

Figure \ref{fig:epsart}
shows the occupation probabilities of the four sites in the system
as a function of time, for the initial conditions indicated in expressions
\ref{eq:one}-\ref{eq:four}. 
Note that for the choice of the potential values corresponding to this
figure, the value of the $\psi(1,1)$ component of the state vector yields
deterministically (i.e. with magnitude either zero or one) the
corresponding states at the right hand side of expressions
\ref{eq:one}-\ref{eq:four}. 

The numerical computation was performed for values of the unitless time
parameter $\tau = t\epsilon/\hbar$ between 0 and 7.665, 
$\alpha/\epsilon=2.350$, and 
the four values of the potential $V(0,0)=-0.003554\epsilon$, 
$V(0,1)=2.124\epsilon$, 
$V(1,0)=2.352\epsilon$, and $V(1,1)=0$. 
The numerical integration of the Schr\"{o}dinger equation was carried
out by factorizing the kinetic and potential energy terms in the exponential
and treating each part exactly\cite{ben}. 
This procedure is correct to second order
in the integration time-step, which was taken to be 
$\Delta\tau=7.665\times 10^{-4}$.
The four final values of $\psi(1,1)$ are
within 0.06, 0.01, 0.04, and 0.04 of their ideal values. Other solutions to the
problem could be found, one needs to adjust the values of $\alpha$, $\tau$,
and the three finite values of the potential until the final value of
$\psi(1,1)$ is within acceptable error. (The fourth value of the potential is 
the arbitrary reference of the potential energy and was chosen as zero.)
The aim at this stage was not to obtain a solution with overwhelming 
accuracy to an idealized model
but to show that nonlinear quantum transformations are available.
A physical realization of such a functional block would necessarily
be more complicated and would require a more careful analysis.

The implementation of the
nonlinear qubit transformation then involves the teleportation of the two
initial qubits into the wavefunction $\psi$, and the teleportation of a single
qubit of information out of $\psi(1,1)$ after a fixed period of time.

In conclusion, we have shown that evolution of a quantum system with a
dynamics controlled by a nonlinear Schr\"{o}dinger equation enables nonlinear
transformations to be carried out on qubits, which may be used to implement
quantum computational algorithms with less restrictions. It may also be
possible to use the nonlinearity to ``saturate" the qubits to some ideal
values close to their initial states, thereby implementing some error
correction. (Quantum state purification through the use of nonlinear 
transformations has been discussed in \cite{pure}.)

A question that
needs to be considered at this point is whether the nonlinear time development
is that of a true quantum system, as microscopic interactions always lead to
unitary development. The assumptions that go into the
development of the approximations that lead to the nonlinear Schr\"{o}dinger
equation may limit the applicability of the corresponding dynamics to the
description of wave phenomena without the quantum
features (such as the second sound effect in superfluid helium). It needs to be
confirmed that the degrees of freedom described by this equation
still maintains the indispensable quantum properties of interference,
entanglement, and the probability interpretation, and is not just a collective 
macroscopic wave phenomena.
Experimental work studying the interference effects in BEC 
systems\cite{BEC-interfere1,BEC-interfere2} and the detailed theoretical
analysis of their entanglement\cite{Shi} seem
to indicate that the Gross-Pitaevskii 
equation may indeed be valid in representing
genuine quantum effects in these systems\cite{note}. Perhaps another way of 
looking at this validity is to
interpret these transformations as high fidelity
representations of nonlinearities through the use of the large number of
quantum background degrees of freedom (individually obeying unitary time
development) contained in the ``mean field" of
the system. 

The author acknowledges helpful discussions with Bilal Tanatar, and partial
support provided by the Turkish Academy of Sciences.

%\begin{references}

%\end{references}

\end{document}